\documentclass[12pt]{article}
\DeclareUnicodeCharacter{2212}{-}

\usepackage{url}
\usepackage{listings,jvlisting}
\usepackage{amssymb}
\usepackage{tabularx}
\usepackage{amsmath}
\usepackage{mdframed}
\usepackage[pdftex]{graphicx}
\usepackage{here}
\usepackage{authblk}
\usepackage{mathtools}
\usepackage{amsthm}
\makeatletter

\@addtoreset{equation}{section}
\makeatother

\makeatletter
\newcommand{\figcaption}[1]{\def\@captype{figure}\caption{#1}}
\newcommand{\tblcaption}[1]{\def\@captype{table}\caption{#1}}
\makeatother

\def\th{{\theta}}

\def\thh{{\hat \th}}

\def\c{{\text{\boldmath $c$}}}
\def\d{{\text{\boldmath $d$}}}

\def\Y{{\text{\boldmath $Y$}}}

\def\Yo{{\overline Y}}

\def\thh{{\hat \th}}

\def\[{{\text{\boldmath $[$}}}
\def\]{{\text{\boldmath $]$}}}

\def\|{{\,|\,}}

\def\/{{\Bigr/\!\!}}

\def\1r{{\rm (1)}}
\def\2r{{\rm (2)}}
\def\3r{{\rm (3)}}
\def\4r{{\rm (4)}}
\def\5r{{\rm (5)}}

\def\Yo{{\overline Y}}
\def\Ybo{{\overline \Y}}

\begin{document}
\title{Adaptive Contrast Test for Dose-Response Studies and Modeling}
\author[1,2]{Masahiro Kojima}
\affil[1]{Kyowa Kirin Co., Ltd}
\affil[2]{The Graduate University for Advanced Studies}

\maketitle

\abstract{We propose a powerful adaptive contrast test with ordinal constraint contrast coefficients determined by observed responses. The adaptive contrast test can perform using easily calculated contrast coefficients and existing statistical software. We provide the sample SAS program codes of analysis and calculation of power for the adaptive contrast test. After the adaptive contrast test shows the statistically significant dose-response, we consider to select the best dose-response model from multiple dose-response models. Based on the best model, we identify a recommended dose. We demonstrate the adaptive contrast test for sample data. In addition, we show the calculation of coefficient, test statistic, and recommended dose for the actual study. We perform the simulation study with eleven scenarios to evaluate the performance of the adaptive contrast test.\\
We confirmed the statistically significant dose-response for the sample data and the actual study. In the simulation study, we confirmed that the adaptive contrast test has higher power in most scenarios compared to the conventional method. In addition, we confirmed that the type 1 error rate of the adaptive contrast test was maintained at a significance level when there was no difference between the treatment groups.\\
We conclude that the adaptive contrast test can be applied unproblematically to the dose-response study.}

\section{Introduction}
\label{sec1}

A primary objective of a dose-response trial is to verify a statistically significant dose-response relationship. After confirming the dose-response, a recommended dose is selected based on efficacy, safety, pharmacokinetic, efficiency of a manufacturing, and so on. In general, various analyses are proposed to confirm the dose-response relationship. For example, there are to identify a recommended dose according to the dose-response from the viewpoint of safety \cite{Liu2015,Yan2017,Lee2019,Lin2020b,Kojima2021a,Kojima2021b}. The analyses method are also proposed to identify a recommended dose from the perspective of safety and efficacy \cite{Mozgunov2020a,Mozgunov2020b,Lin2020a}. In this paper, we consider an analysis method to verify the statistically significant dose-response to verify a proof-of-concept (PoC) in terms of efficacy. Various analyses to confirm dose-response have been proposed \cite{Bretz2005,Pinheiro2014,Ma2020}. In particular, multiple comparison procedures with modeling techniques (MCP-Mod) have been used in various clinical trials \cite{Bornkamp2013,ACZ885H2255,ACZ885M2301,BAF312A2201,BGG492A2207,LCQ908B2302,LCQ908C2201,LIK066A2202,QAW039A2206}. In an MCP part of the MCP-Mod, contrast coefficients are given based on multiple dose-response models, and statistically significant dose-responses are confirmed from contrast tests adjusted for multiplicity. After the dose-response is confirmed, in a Mod part, a dose-response model is selected by using Akaike information criterion (AIC) or Tmax. A recommended dose is selected based on the clinical meaningful difference from the control group. However, we feel a hassle to calculate the contrast coefficient based on the dose-response model and the multivariate t-distribution to adjust for multiplicity. In particular, we can not analyzed the existing statistical analysis software (SAS) procedure. SAS is basically used in a new drug applications. In the R software, there is a MCP-Mod procedure. However, the U.S. Food and Drug Administration (FDA), for example, requires that the MCP-Mod package follow the guidelines of the General Principles of Software Validation. We consider it not easy to confirm to the FDA that the MCP-Mod package is in compliance with the guidelines and that there is no problem using the MCP-Mod package.

In this paper, we propose the simple adaptive contrast test with ordinal constraint contrast coefficients determined by observed response. The adaptive contrast test can perform using easily calculated contrast coefficients and existing statistical software. We provide the sample SAS program codes of analysis and calculation of power for the adaptive contrast test. After the adaptive contrast test shows the statistically significant dose-response, we consider to select the best dose-response model from multiple dose-response models. Based on the best model, we identify a recommended dose. We demonstrate the adaptive contrast test for sample data given by Bretz et al.\cite{Bretz2005}. In addition, we show the calculation of coefficient, test statistic, and recommended dose for the actual study by Akizawa et al. \cite{Akizawa2018}. We perform the simulation study to evaluate the performance of the adaptive contrast test compared to MCP-Mod.

This paper is organized as follows. Chapter 2 introduces the adaptive contrast test and model selection. In addition, the analyses of sample data and actual study are demonstrated. We have shown the configuration of simulation. Chapter 3 describes the results of simulation. Chapter 4 is discussion. Chapter 5 shows a sample program codes for  power calculation and analysis in SAS.

\section{Methods}
\label{sec2}
We consider the randomized, placebo-controlled, multicenter, parallel-group, dose-finding study. The number of arms including the placebo group is $k$. The number of patients treated is $n_i$ $(i=1,2,\ldots,k)$.  The subscript "1" of $n_1$ refers to the placebo group. The observed responses are $\Ybo=(\Yo_1,\ldots,\Yo_k)^T$ such as  the sample means or means adjusted by an analysis of covariance or mixed-effects model for repeated measures. We assume that a larger $\Yo$ indicates a trend toward improvement. However, even if the improvement trend is reversed, an analysis can conduct without any problem. In Section \ref{sec23}, we show an example of improvement as $\Yo$ is lower. The standard deviations are $S_1,\ldots,S_k$. The statistical hypothesis testing for verifying proof of concept (PoC) is conducted by a contrast test. The test statistic is $T=\frac{\sum_{i=1}^{k}c_i\Yo_i}{\sqrt{\left(\sum_{i=1}^k \frac{c_i^2}{n_i}\right)S^2}}$, where $\sum_{i=1}^k c_i=0$ and $S^2=\frac{1}{\sum_{i=1}^k n_i - k}\sum_{i=1}^k \left(n_i-1\right)S_i^2$. When $T$ exceeds the upper 2.5\% point of the distribution followed by the statistic, a statistically significant dose-response is shown, and the PoC for an investigational drug is accepted.

\subsection{Adaptive contrast test}\label{sec21}
\begin{figure}[h]
  \begin{center}
  \includegraphics[width=15cm]{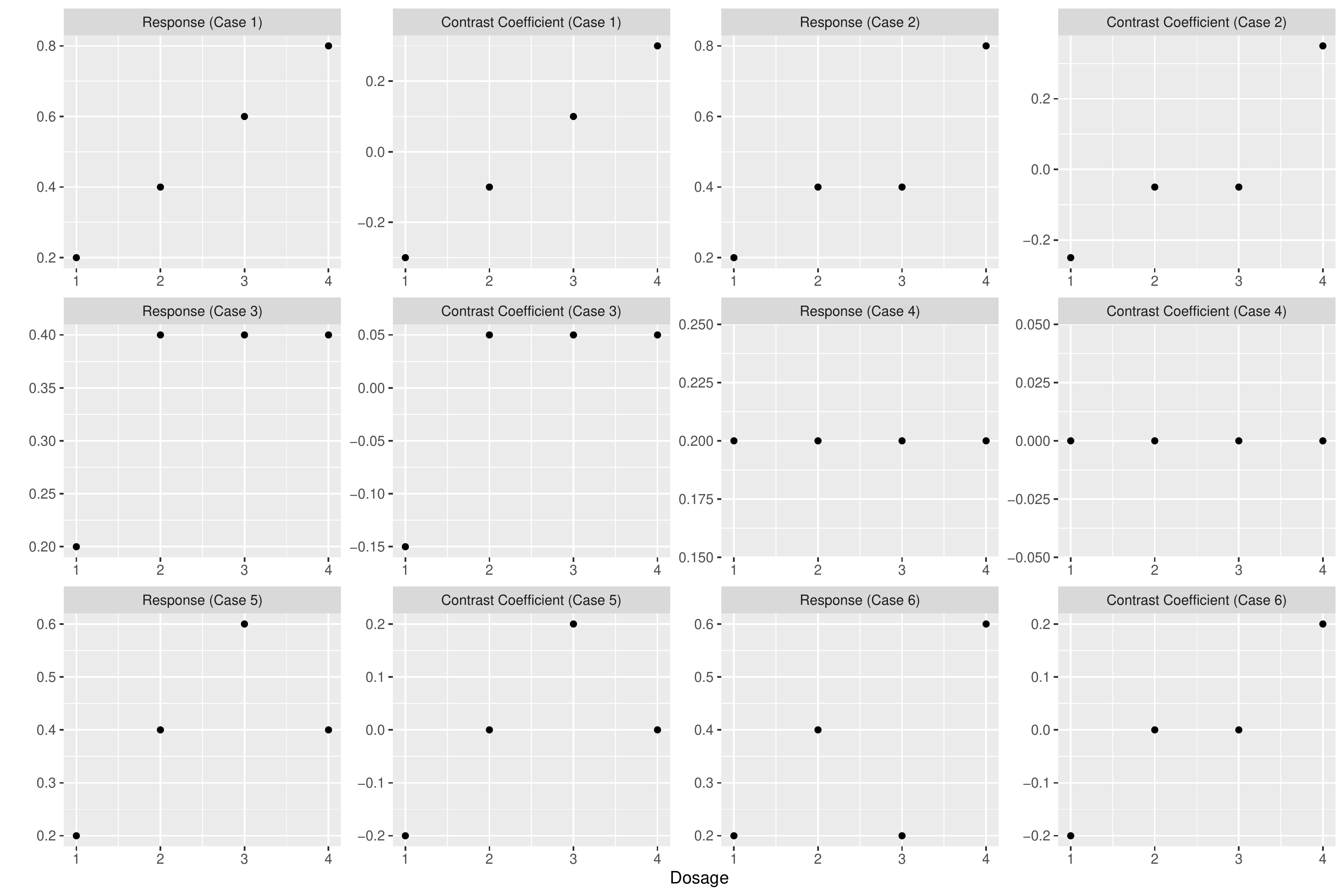}
  \caption{Examples of actual responses $\Ybo$ and  coefficients $\c$}
  \label{Figure1}
  \end{center}
\end{figure}

We propose a novel adaptive contrast test. First of all, we give an ordinal constraint of each element of the contrast coefficients $\c$. For example, we assume that $\c$ increases quasi-monotonically in a dose-dependent, the ordinal constraint of $\c$ is $c_1\leq c_2\leq c_3\leq c_4\leq \cdots \leq c_k$. The constraint should be defined before the start of study. Under the constraint, the each of $\c$ is calculated based on the observed responses $\Ybo$. $c_1$ is given by $\frac{1}{k}\left((k-1)\Yo_1-\sum^k_{i=2}\underset{j\in {1,\ldots,i}}{\operatorname{max}}(\Yo_j)\right)$ and $c_i=\underset{j\in {1,\ldots,i}}{\operatorname{max}}(\Yo_j)-\underset{j\in {1,\ldots,i-1}}{\operatorname{max}}(\Yo_j)+c_{i-1}$, $i=2,3,\ldots,k$. The reason for taking the maximum value is to satisfy the ordinal constraint $c_i\leq c_j$ at a dose $j$ that is larger than dose $i$. We show examples of observed response $\Ybo$ and contrast coefficient $\c$ for four arms in Figure \ref{Figure1}. The constraint of $\c$ is $c_1\leq  c_2\leq c_3$. The $c_4$ has no constraint because we are interested in the $\c$ adapting flexibly to an umbrella shape. For the case 1 to case 5, the $\c$ shows a similar trend in the observed response. For the case 6, the observed response of dose $3$ is lower than that of dose 2. Hence, the contrast coefficient of dose $2$ is the same with the coefficient of dose $3$. The formulas for each $\c$ in the example are $c_1=\frac{1}{4}\left(3\Yo_1-\underset{j\in {1,2}}{\operatorname{max}}(\Yo_j)-\underset{j\in {1,2,3}}{\operatorname{max}}(\Yo_j)-\Yo_4\right)$, $c_2=\underset{j\in {1,2}}{\operatorname{max}}(\Yo_j)-\Yo_1+c_1$, $c_3=\underset{j\in {1,2,3}}{\operatorname{max}}(\Yo_j)-\underset{j\in {1,2}}{\operatorname{max}}(\Yo_j)+c_2$, $c_4=\Yo_4-\underset{j\in {1,2,3}}{\operatorname{max}}(\Yo_j)+c_3$. As an example of the specific calculation of $\c$ using actual response values, in the case 1, for $\Ybo=(0.2,0.4,0.6,0.8)$, each of $\c$ is $c_1=\frac{1}{4}(3\times 0.2-0.4-0.6-0.8)=-0.3$, $c_2=0.4-0.2+(-0.3)=-0.1$, $c_3=0.6-0.4+(-0.1)=0.1$, and $c_4=0.8-0.6+0.1=0.3$. In the case 6, under $\Ybo=(0.2,0.4,{\bf 0.2},0.6)$, each of $\c$ is $c_1=\frac{1}{4}(3\times 0.2-0.4-{\bf 0.4}-0.6)=-0.2$, $c_2=0.4-0.2+(-0.2)=0$, $c_3={\bf 0.4}-0.4+0=0$, and $c_4=0.6-{\bf 0.4}+0=0.2$. The test statistic $T$ is calculated using the calculated $\c$. Because the test statistic using $\c$ with ordinal constraints does not follow the t-distribution, we use the permutation method to calculate the p-value. The permutation method is design-based analysis method which is suitable for randomized design in dose-response studies. In other words, the randomized design is not a random sampling design.  If all response values are the same or all the investigational drug groups are lower than the placebo group, the test statistic is set to zero.

\subsubsection{Modeling}\label{sec211}
When a statistically significant dose-response is verified, we are interested in identifying the recommended dose from a dose-response model fitting the observed response. We consider to use the AIC to select a dose-response model. Candidate models include Linear, Log-Linear, Emax, Exponential, Quadratic, and Logistic models. If the best model is selected, the recommended dose should be selected using minimal effective dose (MED). The MED is a clinically meaningful difference from placebo. If there is a clinically meaningful change in response from baseline in the medical guideline, the recommended dose can select from the doses that are changed meaningful rather than looking at the difference from placebo.

\subsection{Analysis of sample dataset}\label{sec22}
We demonstrate the adaptive contrast test by using the sample dataset given by Bretz et al.\cite{Bretz2005}. The sample dataset consists of data from 20 patients per group in the placebo and four drug groups (dosages: 0.05, 0.20, 0.60, and 1) in a randomized trial. The responses of each group follow a normal distribution. The sample means are $\Ybo=(0.345,0.457,0.810,0.934,0.949)^T$, the standard deviations are $S_1=0.517$, $S_2=0.490$, $S_3=0.740$, $S_4=0.765$, and $S_5=0.947$. The elements of $\c$ are $c_1=\frac{1}{5}(4\times 0.345-0.457-0.810-0.934-0.949)=-0.354$, $c_2=0.457-0.345+c_1=-0.242$, $c_3=0.810-0.457+c_2=0.111$, $c_4=0.934-0.810+c_3=0.235$, and $c_5=0.949-0.934+c_4=0.250$. The test statistic is $T=3.330$, the one-sided p-value of the permutation method is $p=0.0003$. We can have confirmed the statistically significant dose-response.

We consider the model selection. We choose the best model from Emax, Linear log, Linear, Exponential, Quadratic, and Logistic models in terms of prediction for each dose response by using the AIC. Because the AIC of Emax model is the smallest, Emax is selected as the dose-response model. We have summarized the transition for each model in Figure 2.

\begin{figure}[h]
  \begin{center}
  \includegraphics[width=12cm]{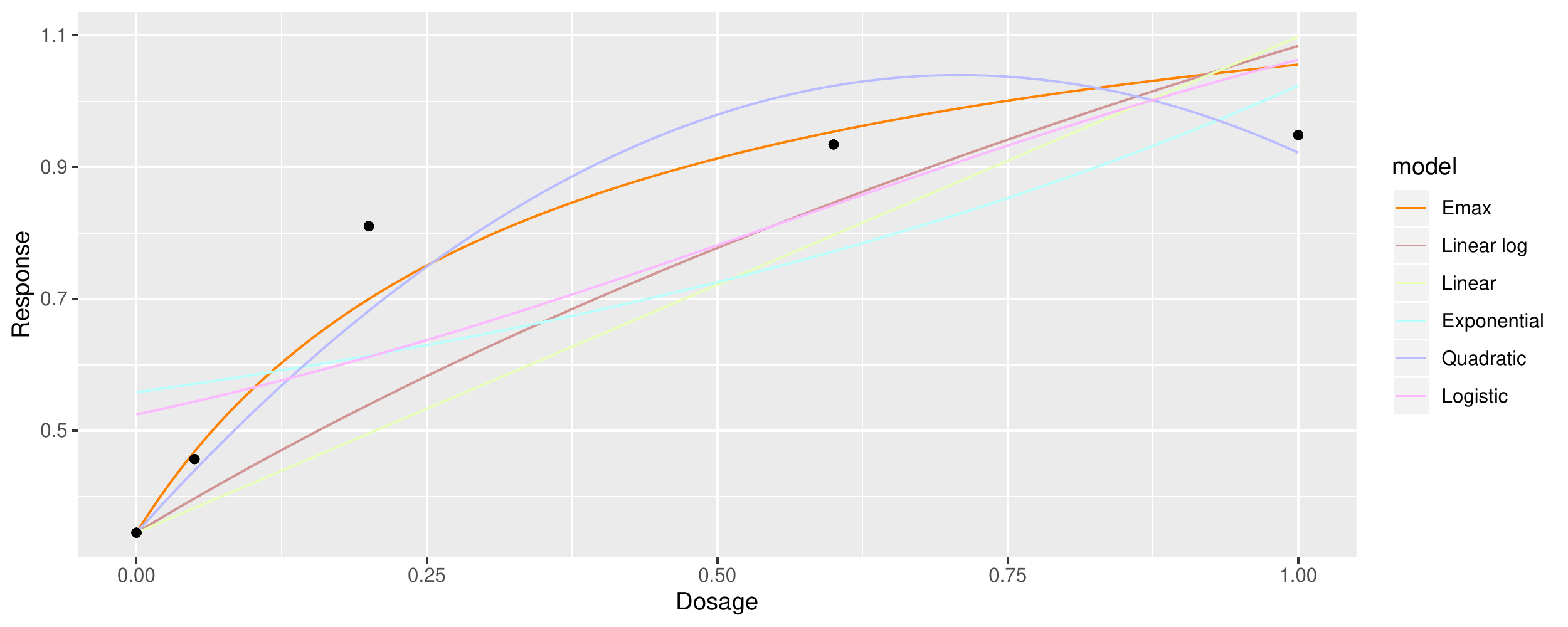}
  \caption{Dose-response model and observe responses shown as dots}
  \label{Figure2}
  \end{center}
\end{figure}

\subsection{Actual study (Phase 2b study of evocalcet)}\label{sec23}
We re-analyze the phase 2b study of evocalcet for hemodialysis patients with secondary hyperparathyroidism using the summary data. The objective of the study is to confirm the PoC of efficacy for the randomized, double-blind, placebo-controlled, multicenter, parallel-group, dose-finding design. The patients were assigned randomly to a placebo, 0.5, 1, 2 mg/day  of evocalcet group for 3 weeks treatment period. The primary endpoint is the percent change from baseline in intact parathyroid hormone (PTH) at the end of treatment. The primary analysis is contrast test with seven contrast patterns for a dose-response. The PoC was shown by a statistically significant decrease in the percent change in iPTH. The secondary analysis calculated the sample mean and standard deviation of percent change from baseline in the intact PTH of each group at end of treatment. The results (Mean$\pm$SD) of percent change from baseline in the intact PTH were 5.44$\pm$25.85\% in placebo, −8.40$\pm$25.43\% in 0.5 mg, −10.56$\pm$22.86\% in 1 mg, and −20.16$\pm$34.23\% in 2 mg. Because a lower value for the percent change indicates an clinical improvement, a constraint on the contrast coefficient is given as $c_1\geq c_2\geq c_3\geq c_4$. By calculating coefficients $\c$ based on the formula replacing the maximum function with a minimum function, $c_1=13.86$, $c_2=0.02$, $c_3=-2.14$, and $c_4=-7.56$. Pooled variance is $S^2=773.17$. $T=\frac{13.86*5.44-0.02*8.40+2.14*10.56+11.74*20.16}{\sqrt{S^2*(13.86^2/28+0.02^2/30+(-2.14)^2/30+(-11.74)^2/28)}}=3.54$. Because we can not have access to the individual data for this study, we show the upper points for t-distribution. The upper $2.5\%$ point of the t-distribution is $1.98$, The upper $0.25\%$ point of the t-distribution is $2.86$. The upper $0.05\%$ point of the t-distribution is $3.38$. We have confirmed that the result is statistically significant even when the significance level is sufficiently small. We confirmed the power based on the sample mean and standard deviation in this study, and the power was 92.04\%. Based on these results, we assume that the permutation method shows statistically significant. Although this study shows the 90.0\% power via multiple contrasts test, the power for the adaptive contrast test following the setting of sample size in this study was 91.7\%.

We consider the model selection. E$_{max}$ model is $E_0+E_{max}d/(\th+d)$. $E_0$ is initial value $5.44$ at placebo. E$_{max}$ is the minimum value $-20.16$ at 2 mg, and $\th$ is parameter. The estimator $\thh$ is $0.40$. The AIC is $22.4$.  Linear log-dose model is $E_0+\th\log(d+1)$. The estimator $\thh$ is $-24.21$. The AIC is $21.3$. Linear model is $E_0+\th d$. The estimator $\thh$ is -14.12. The AIC is 25.9. Exponential model is $E_0+\th_1\exp(d/\th_2)$. $\th_1$ and $\th_2$ is parameter. The estimators $\thh_1$ and $\thh_2$ are $1.48$ and $-6.87$, respectively. The AIC is 35.1. Quadratic model is $E_0+\th_1 d+\th_2 d^2$. The estimator $\thh_1$ is -24.19 and $\thh_2$ is $5.79$. The AIC is $22.9$. Logistic model is $E_0+E_{max}/(1+\exp((\th_1-d)/\th_2))$. The estimator $\thh_1$ is 0.66 and $\thh_2$ is $0.36$. The AIC is $25.8$. The minimum AIC is shown for the Linear log-dose model, we select the Linear log-dose model as the best dose-response model. We show the all models in Figure 3. We show examples of recommended dose selection. If a 10\% decrease in the rate of change in iPTH has clinical implications, we can select the dose 1.0 or more. If a difference of 10\% or more from placebo is a clinical meaningful, we can select the dose 0.5 or more.

\begin{figure}[h]
  \begin{center}
  \includegraphics[width=12cm]{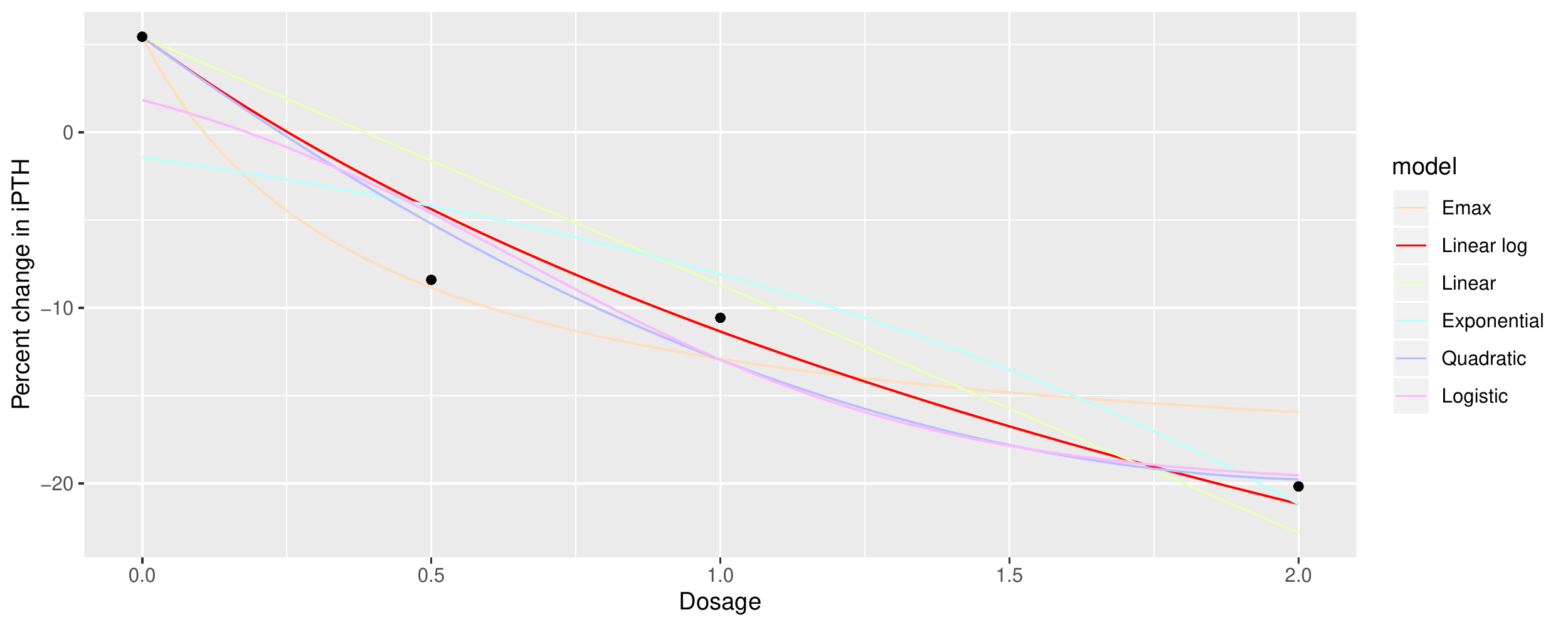}
  \caption{Figure 3. Model}
  \label{Figure3}
  \end{center}
\end{figure}

\subsection{Simulation study}
We evaluate the statistical power of adaptive contrast test compared to the MCP-Mod via simulation study. We assume the randomized dose-response study with five arms and one-sided significance level 2.5\%. The dosages are $\d=(d_1,d_2,d_3,d_4,d_5)^T=(0,0.05,0.2,0.6,1.0)^T$. The number of simulations was set to 10,000. The constraint of contrast coefficient is $c_1\leq c_2\leq c_3\leq c_4$. The coefficient $c_5$ for the highest dose has no constraint because we also consider the umbrella shape. The number of permutations for permutation method is 50,000. The MCP part of MCP-Mod evaluates the models shown in the Table \ref{Table1} referred by \cite{Bretz2005}. The true mean value of each dose is shown in Table \ref{Table2}. The scenario 1 refers to the constant mean values to confirm the significance level maintaining at 2.5\%. For the scenario 2 to the scenario 7, the true mean values are generated by the dose-response models in the Table \ref{Table1}. For the scenario 8 to the scenario 11, we assume the results with the dose-response relationship that is not based on a dose-response model. The standard deviation is 1.5. 

\begin{table}[h!]
  \begin{center}
\caption{Dose-response model\label{Table1}}
{\tabcolsep=4.25pt
\begin{tabular}{|c|c|}\hline
Model name & Equation\\\hline
Constant & $0.2$  \\
Linear &  $0.2+0.6d_i$ \\
Linear in log-dose &  $0.2+0.6\log(5d_i+1)/\log(6)$ \\
Emax &  $0.2+0.7d_i/(0.2+d_i)$ \\
Exponential &  $0.183+0.017\exp(2d_i\log(6))$ \\
Quadratic &  $0.2+2.049d_i-1.749d_i^2$ \\
Logistic &  $0.193+0.607/(1+\exp(10\log(3)(0.4-d_i)))$ \\\hline
\end{tabular}}
  \end{center}
\end{table}

\begin{table}[h!]
  \begin{center}
\caption{True mean value\label{Table2}}
{\tabcolsep=4.25pt
\begin{tabular}{|c|c|}\hline
 & true mean values\\\hline
Scenario 1 (Constant) & $(0.2,0.2,0.2,0.2,0.2)$  \\
Scenario 2 (Linear) &  $(0.2,0.23,0.32,0.56,0.8)$ \\
Scenario 3 (Linear in log-dose) &  $(0.2,0.275,0.432,0.664,0.8)$ \\
Scenario 4 (Emax) &  $(0.2,0.34,0.55,0.725,0.783)$ \\
Scenario 5 (Exponential) &  $(0.2,0.201,0.206,0.226,0.264)$ \\
Scenario 6 (Quadratic) &  $(0.2,0.298,0.54,0.8,0.5)$ \\
Scenario 7 (Logistic) &  $(0.271,0.289,0.362,0.631,0.767)$ \\
Scenario 8 &  $(0.2,0.4,0.6,0.6,0.8)$ \\
Scenario 9 &  $(0.2,0.4,0.6,0.6,0.6)$ \\
Scenario 10 &  $(0.2,0.6,0.6,0.6,0.6)$ \\
Scenario 11 &  $(0.2,0.6,0.6,0.8,0.8)$ \\\hline
\end{tabular}}
  \end{center}
\end{table}

\section{Results}\label{sec3}
\begin{figure}[h]
  \begin{center}
  \includegraphics[width=12cm]{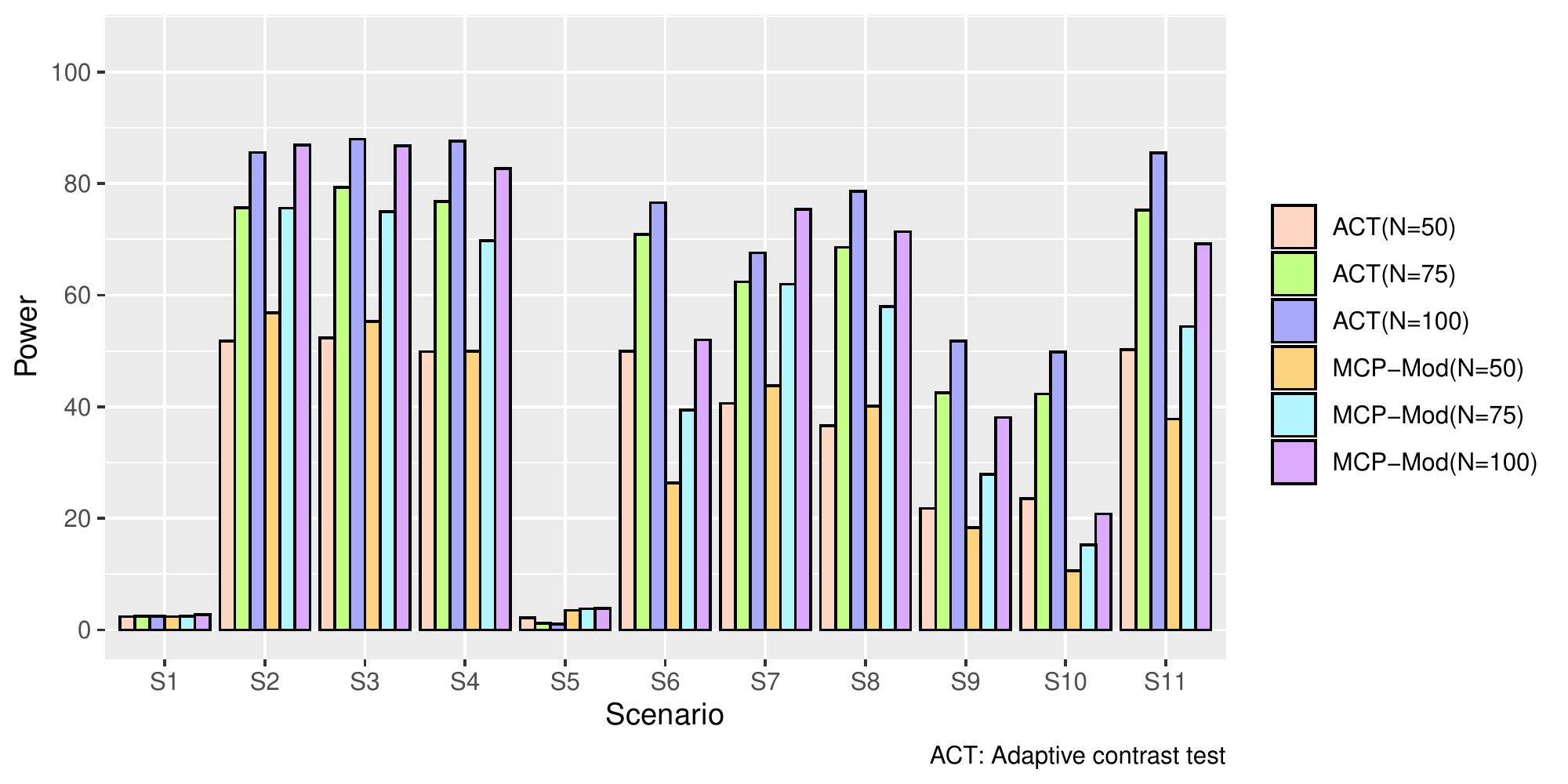}
  \caption{Power of each scenario in simulation study}
  \label{Figure4}
  \end{center}
\end{figure}

The results of the simulation study are shown in Table \ref{Table3} and Figure \ref{Figure4}. The scenario 1 confirmed that the significance level of 2.5\% was maintained for all adaptive contrast test and MCP-Mod. The power increased with increasing sample size for the adaptive contrast test and MCP-Mod except in Scenarios 1 and 5. For the adaptive contrast test (N=100), the power was higher than the MCP-Mod in scenarios 3, 4, 6, 8, 9, 10, and 11, while MCP had higher power in scenarios 2, 5, and 7. MCP had lower power in scenarios 9, 10, and 11, which were not generated from the model, compared to scenarios 2 to 7, which were generated from the model. Supplementally, we show the results of quasi-monotonically increasing for the ordinal constraint of all contrast coefficients (not apply Umbrella shape) in Supplementary Analysis \ref{sec6}.

\begin{table}[h]
  \begin{center}
\caption{Power of each scenario in simulation study\label{Table3}}
{\tabcolsep=4.25pt
\begin{tabular}{|c|c|c|c|c|c|c|}\hline
& \multicolumn{3}{c|}{Adaptive contrast test} & \multicolumn{3}{c|}{MCP-Mod}\\\hline
 & $N=50$ & $N=75$ & $N=100$ & $N=50$ & $N=75$ & $N=100$\\ \hline
Scenario 1 &  2.40&2.45&2.52&2.39&2.46&2.66\\
Scenario 2 &  51.82&75.71&85.63&56.81&75.65&86.86\\
Scenario 3 &  52.26&79.32&88.01&55.27&75.04&86.81\\
Scenario 4 &  49.92&76.84&87.61&49.97&69.79&82.69\\
Scenario 5 &  2.20&1.15&1.06&3.45&3.80&3.85\\
Scenario 6 &  50.01&70.85&76.57&26.37&39.42&51.97\\
Scenario 7 &  40.57&62.43&67.57&43.81&61.97&75.41\\
Scenario 8 &  36.59&68.59&78.57&40.12&57.98&71.42\\
Scenario 9 &  21.80&42.50&51.79&18.27&27.92&38.06\\
Scenario 10 &  23.53&42.27&49.80&10.61&15.30&20.84\\
Scenario 11 & 50.17&75.31&85.46&37.82&54.42&69.23\\\hline
\end{tabular}}
  \end{center}
\end{table}

\section{Discussion}\label{sec4}
We proposed the adaptive contrast test and model selection. The contrast coefficients are given the ordinal constraint before the study starts and adaptively determined in a data-dependent. We have confirmed that the adaptive contrast test has higher power because of the contrast coefficients determined adaptively. On the practical side, the contrast coefficients are easy to calculate. The statistical test is performed by permutation method, which can be easily computed using, for example, the multtest procedure in SAS, permute in STATA, or perm package in R. In addition, we provide the sample SAS program codes of analysis and calculation of power for the adaptive contrast test in Chapter \ref{sec5}. We proposed a procedure to choose a dose-response model from candidate models and select a recommended dose after a statistically significant dose-response has been confirmed. We also provide the sample program codes using the existing SAS procedure for model selection in Chapter \ref{sec5}.	

We demonstrated the adaptive contrast test for the sample study given by Bretz et al.\cite{Bretz2005}. We have confirmed the statistically significant dose-response. In addition, we selected the dose-response model via the AIC. In addition, We re-analyzed the actual phase 2b study by Akizawa et al.\cite{Akizawa2018}. We showed the determined contrast coefficients and the contrast test statistic. The test statistic implied that the results were statistically significant, and we selected the best dose-response model. We presented recommended doses with clinically meaningful efficacy based on the dose-response models. The power of the adaptive contrast test showed higher than the permutation test for multiple contrasts used in the actual phase 2b study.

We performed the simulation study to evaluate the power of adaptive contrast test compared to the MCP-Mod. In many scenarios, the adaptive contrast test has higher power than the MCP-Mod. We consider that the power was high by identifying the optimal contrast coefficients in a data-dependent. We confirmed that the one-sided type 1 error rate of the adaptive contrast test was maintained at 2.5\% when there was no difference between the treatment groups. Hence, there was no problem with the performance. The power of MCP-Mod was relatively high when the true mean of each group was based on the dose-response model. However, the power of MCP-Mod decreased when the true mean was generated not-dose-response model. In reality, because the true mean values do not transition based on the dose-response model the MCP-Mod may not be able to maintain the expected power. For the ordinal constraint of the contrast coefficient, when the quasi-monotonic increase assumption was made for all coefficients without assuming an umbrella type, the power increased for the all scenarios except for the scenario generated from an umbrella type. We recommend the assumption of quasi-monotonic increase for all coefficients when no umbrella type is assumed in the efficacy data.

The adaptive contrast test is a powerful test that can perform using easily calculated contrast coefficients and existing statistical software. We confirmed that the adaptive contrast test is the higher power than not only the permutation test with multiple contrast patterns but also the MCP-Mod. When we plan to use the permutation test with multiple contrast patterns and MCP-Mod, we need to explain the procedure of those methods, assumption of dose-response models and adjustment of multiplicity to clinicians and decision-makers. We proposed the analysis method that can avoid the multiplicity and be easy-to-understand of analysis procedure. The adaptive contrast test can be easy to execute by the simple analysis program. We provide the sample SAS codes, SAS is basically used in new drug applications. Therefore, we hope that the adaptive contrast test will be used in many dose-response studies.

\section{Software}\label{sec5}

\begin{lstlisting}[caption=Sample program code of analysis of biom dataset]
proc import out=dat
	datafile="\biom.xlsx"
	/* Add path. biom.xlsx converted from data(biom) of R MCPMoD package*/
	dbms=Excel replace; 
	getnames=no; 
run;

/*Variable name tn is arm name (numeric) and res is response (numeric)*/

data dat;
	length t $200.;
	set dat;
	if tn=0 then t="1_0";
	else if tn=0.05 then t="2_0.05";
	else if tn=0.2 then t="3_0.2";
	else if tn=0.6 then t="4_0.6";
	else if tn=1 then t="5_1";
run;

proc univariate data=dat noprint;
	where tn=0;
	var res;
	output out=out1 mean=mean1 std=std1;
run;

proc univariate data=dat noprint;
	where tn=0.05;
	var res;
	output out=out2 mean=mean2 std=std2;
run;
proc univariate data=dat noprint;
	where tn=0.2;
	var res;
	output out=out3 mean=mean3 std=std3;
run;
proc univariate data=dat noprint;
	where tn=0.6;
	var res;
	output out=out4 mean=mean4 std=std4;
run;
proc univariate data=dat noprint;
	where tn=1;
	var res;
	output out=out5 mean=mean5 std=std5;
run;

data out;
	merge out1-out5;
run;

%macro _do;
data out;
	set out;
	mean1=round(mean1,1E-5);
	mean2=round(mean2,1E-5);
	mean3=round(mean3,1E-5);
	mean4=round(mean4,1E-5);
	mean5=round(mean5,1E-5);
	max2=max(of mean1-mean2);
	max3=max(of mean1-mean3);
	max4=max(of mean1-mean4);
	max5=max(of mean1-mean5);
	c1=round(-(max2+max3+max4+max5-4*mean1)/5,1E-5);
	c2=round((max2-mean1)+c1,1E-5);
	c3=round((max3-max2)+c2,1E-5);
	c4=round((max4-max3)+c3,1E-5);
	c5=round((max5-max4)+c4,1E-5);
	call symput("cc1", c1);
	call symput("cc2", c2);
	call symput("cc3", c3);
	call symput("cc4", c4);
	call symput("cc5", c5);
	if c1=0 and c2=0 and c3=0 and c4=0 and c5=0 then do; %let _FL=Y; end;
	else do;
		%let _FL=N;
	end;

run;

%if &_FL.=Y %then %do;
  data pValues_1;
  	set pValues_1;
	Permutation=1;
  run;
 %end;
 %else %do;
  ods output pValues = pValues_1;

  proc multtest data=dat permutation nsample=10000 seed=2021;
    class t;
    test mean (res / ddfm=pooled upper);
      contrast 'Adaptive Contrast' &cc1. &cc2. &cc3. &cc4. &cc5.;
  run;
  ods listing;
  %end;

  data pValues_1;
  	set pValues_1;
	c1=&cc1.;
	c2=&cc2.;
	c3=&cc3.;
	c4=&cc4.;
	c5=&cc5.;
  run;
  %mend;

%_do;
/*Dataset pValues_1 shows p-value.*/
  
 /*AIC is derived below codes*/
data dr;
input d res E0 Emax;
datalines;
0 0.34491 0.34491 0.94871
0.05 0.45675 0.34491 0.94871
0.2 0.81032 0.34491 0.94871
0.6 0.93444 0.34491 0.94871
1 0.94871 0.34491 0.94871
;
run;
title "Emax model" ;
ods output FitStatistics=AIC_Emax ;
proc nlmixed data = dr;
  parms ED50 = 1 SD=1;
  mu = E0 +Emax*d / (ED50+d);
  model res ~ normal(mu, SD**2);
run;
ods output close;

title "Linear log-dose model" ;
ods output FitStatistics=AIC_Lld ;
proc nlmixed data = dr;
  parms de = 1 SD=1;
  mu = E0 +de*log(d+1);
  model res ~ normal(mu, SD**2);
run;
ods output close;

title "Linear model" ;
ods output FitStatistics=AIC_L ;
proc nlmixed data = dr;
  parms de = 1 SD=1;
  mu = E0 +de*d;
  model res ~ normal(mu, SD**2);
run;
ods output close;

title "Exponential model" ;
ods output FitStatistics=AIC_Exp ;
proc nlmixed data = dr;
  parms sl=1 de = 1 SD=1;
  mu = E0 +sl*exp(d/de);
  model res ~ normal(mu, SD**2);
run;
ods output close;

title "Quadratic model" ;
ods output FitStatistics=AIC_Q ;
proc nlmixed data = dr;
  parms be1 = 1 be2=1 SD=1;
  mu = E0 +be1*d+be2*d**2;
  model res ~ normal(mu, SD**2);
run;
ods output close;

title "Logistic model" ;
ods output FitStatistics=AIC_Log ;
proc nlmixed data = dr;
  parms ED50=1 de=1  SD=1;
  mu = E0 +Emax/(1+exp((ED50-d)/de));
  model res ~ normal(mu, SD**2);
run;
ods output close;

\end{lstlisting}

\begin{lstlisting}[caption=Sample program code for calculation of power for the adaptive contrast test]
data res1; set _NULL_; run;
%macro _func(_num,_m1,_m2,_m3,_m4,_m5,_sd);
data test;
	CALL STREAMINIT(100);
	do i=1 to 10000;
		do j=1 to &_num.;
			x1=rand('NORMAL',&_m1.,&_sd.);
			x2=rand('NORMAL',&_m2.,&_sd.);
			x3=rand('NORMAL',&_m3.,&_sd.);
			x4=rand('NORMAL',&_m4.,&_sd.);
			x5=rand('NORMAL',&_m5.,&_sd.);
			output;
		end;
	end;
run;

proc univariate data=test noprint;
	by i;
	var x1;
	output out=out1 mean=mean1 std=std1;
run;

proc univariate data=test noprint;
	by i;
	var x2;
	output out=out2 mean=mean2 std=std2;
run;

proc univariate data=test noprint;
	by i;
	var x3;
	output out=out3 mean=mean3 std=std3;
run;

proc univariate data=test noprint;
	by i;
	var x4;
	output out=out4 mean=mean4 std=std4;
run;

proc univariate data=test noprint;
	by i;
	var x5;
	output out=out5 mean=mean5 std=std5;
run;


data out;
	merge out1-out5;
	by i;
run;

data out;
	set out;
	mean1=round(mean1,1E-5);
	mean2=round(mean2,1E-5);
	mean3=round(mean3,1E-5);
	mean4=round(mean4,1E-5);
	mean5=round(mean5,1E-5);
	max2=max(of mean1-mean2);
	max3=max(of mean1-mean3);
	max4=max(of mean1-mean4);
	max5=max(of mean1-mean5);
	c1=round(-(max2+max3+max4+mean5-4*mean1)/5,1E-5);/
		*Non-Umblella, round(-(max2+max3+max4+max5-4*mean1)/5,1E-5);*/
	c2=round((max2-mean1)+c1,1E-5);
	c3=round((max3-max2)+c2,1E-5);
	c4=round((max4-max3)+c3,1E-5);
	c5=round((mean5-max4)+c4,1E-5);/*Non-Umblella, round((max5-max4)+c4,1E-5);*/
	call symput("cc1", c1);
	call symput("cc2", c2);
	call symput("cc3", c3);
	call symput("cc4", c4);
	call symput("cc5", c5);
	S=(std1**2+std2**2+std3**2+std4**2)*(&_num.-1)/(&_num.*4-4);
	if c1=0 and c2=0 and c3=0 and c4=0 and c5=0 then do; %let _FL=Y; end;
	else do;
		%let _FL=N;
	end;
run;

*proc freq data=out noprint;
*	table FL/out=res;
*run;

data test1(keep=val trtpn i);
	set test;
	rename x1=val;
	TRTPN=1;
run;
data test2(keep=val trtpn i);
	set test;
	rename x2=val;
	TRTPN=2;
run;
data test3(keep=val trtpn i);
	set test;
	rename x3=val;
	TRTPN=3;
run;
data test4(keep=val trtpn i);
	set test;
	rename x4=val;
	TRTPN=4;
run;
data test5(keep=val trtpn i);
	set test;
	rename x5=val;
	TRTPN=5;
run;

data testt;
	set test1-test5;
run;
%if &_FL.=Y %then %do;
  data pValues_1;
  	set pValues_1;
	Permutation=1;
  run;
 %end;
 %else %do;
  ods output pValues = pValues_1;

  proc sort data=testt;by i ;run;
  proc multtest data=testt permutation nsample=50000 seed=2021;
  by i;
    class TRTPN;
    test mean (val / ddfm=pooled upper);
      contrast 'Adaptive Contrast' &cc1. &cc2. &cc3. &cc4. &cc5.;
  run;
  ods listing;
  %end;
  
  data pValues_1;
  	set pValues_1;
	if Permutation<0.025 then FL="Y";
	else if Permutation>=0.025 then FL="N";
  run;

proc freq data=pValues_1 noprint;
	table FL/out=res2;
run;

data res2;
	set res2;
  	SS=&_num.;
  	m1=&_m1.;
  	m2=&_m2.;
  	m3=&_m3.;
  	m4=&_m4.;
  	m5=&_m5.;
  	sd=&_sd.;
run;	

  data res1;
  	set res1 res2;
run;

%mend _func;

%macro _loop(_n);
	%_func(&_n.,0.2,0.2,0.2,0.2,0.2,1.5);
	%_func(&_n.,0.2,0.23,0.32,0.56,0.8,1.5);
	%_func(&_n.,0.2,0.275,0.432,0.664,0.8,1.5);
	%_func(&_n.,0.2,0.34,0.55,0.725,0.783,1.5);
	%_func(&_n.,0.2,0.201,0.206,0.226,0.264,1.5);
	%_func(&_n.,0.2,0.298,0.54,0.8,0.5,1.5);
	%_func(&_n.,0.271,0.289,0.362,0.631,0.767,1.5);
	%_func(&_n.,0.2,0.4,0.6,0.6,0.8,1.5);
	%_func(&_n.,0.2,0.4,0.6,0.6,0.6,1.5);
	%_func(&_n.,0.2,0.6,0.6,0.6,0.6,1.5);
	%_func(&_n.,0.2,0.6,0.6,0.8,0.8,1.5);
%mend;

%_loop(50);
%_loop(75);
%_loop(100);

proc export data = res1
	outfile = "output.xlsx" /*Add path*/
dbms = xlsx replace;
run;

\end{lstlisting}
\section{Supplementary Result}\label{sec6}
We show supplemental result with the ordinal constraint with assuming an umbrella type (Adaptive contrast test 1) and without assuming an umbrella type (Adaptive contrast test 2). The adaptive contrast test 1 assumes $c_1\leq c_2\leq c_3\leq c_4$. The adaptive contrast test 2 assumes $c_1\leq c_2\leq c_3\leq c_4\leq c_5$. The adaptive contrast test 2 was higher power than the adaptive contrast test 1 except for the scenario 6 (Umbrella shape).

\begin{table}[h]
  \begin{center}
\caption{Power of each scenario in supplemental simulation study\label{Table3}}
{\tabcolsep=4.25pt
\begin{tabular}{|c|c|c|c|c|c|c|}\hline
 & \multicolumn{3}{c|}{Adaptive contrast test 1} & \multicolumn{3}{c|}{Adaptive contrast test 2}\\\hline
 & $N=50$ & $N=75$ & $N=100$ & $N=50$ & $N=75$ & $N=100$\\ \hline
Scenario 1 &  2.40&2.45&2.52&2.47&2.45&2.26\\
Scenario 2 &  51.82&75.71&85.63&61.32&83.42&88.47\\
Scenario 3 &  52.26&79.32&88.01&62.53&83.31&88.06\\
Scenario 4 &  49.92&76.84&87.61&62.94&80.2&85.27\\
Scenario 5 &  2.20&1.15&1.06&3.06&4.59&3.36\\
Scenario 6 &  50.01&70.85&76.57&48.37&64.76&72.12\\
Scenario 7 &  40.57&62.43&67.57&49.86&72.44&77.18\\
Scenario 8 &  36.59&68.59&78.57&51.73&72.62&80.3\\
Scenario 9 &  21.80&42.50&51.79&38.07&50.43&54.88\\
Scenario 10 & 23.53&42.27&49.80&39.27&50.64&66.03\\
Scenario 11 & 50.17&75.31&85.46&63.28&79.6&86.49 \\\hline
\end{tabular}}
  \end{center}
\end{table}

\bigskip
\noindent
{\bf Acknowledgments.}\ \ 
The author would like to thank Associate Professor Hisashi Noma for his encouragement and helpful suggestions.





\nocite{*}
\bibliography{manuscript}%



\end{document}